\documentclass[twocolumn,showpacs,pre,aps,amssymb,amsmath]{revtex4}
\usepackage{graphicx}
\usepackage{amssymb}
\usepackage{amsmath}
\renewcommand{\ge}{\geqslant}

\renewcommand{\epsilon}{\varepsilon}
\begin{document}
\title{\bf Hamiltonian fractals and chaotic scattering of passive particles by a
topographical vortex and an alternating current}
\author{M. Budyansky, M. Uleysky, S. Prants}
\affiliation{Laboratory of Nonlinear Dynamical Systems,
V.I.Il'ichev Pacific Oceanological Institute of the Russian Academy of Sciences,
690041 Vladivostok, Russia}
\date{\today}
\begin{abstract}
We investigate the dynamics of passive particles  in a two-dimensional
incompressible open flow composed of a fixed topographical point vortex and a
background current with a periodic component. The tracer dynamics is found
to be typically chaotic in a mixing region and regular in far upstream
and downstream regions of the flow. Chaotic advection of tracers is proven
to be of a homoclinic nature with transversal intersections of stable and
unstable manifolds of the saddle point. In spite of simplicity of the
flow, chaotic trajectories are very complicated alternatively sticking nearby
boundaries of the vortex core and islands of regular motion and wandering
in the mixing region. The boundaries act as dynamical traps for advected
particles with a broad distribution of trapping times. This implies
the appearance of fractal-like scattering function: dependence of
the trapping time on initial positions of the tracers. It is confirmed
numerically by computing a trapping map and trapping time distribution
which is found to be initially Poissonian with a crossover to a power law
at the PDF tail. The mechanism of generating the fractal is shown
to resemble that of the Cantor set with the Hausdorff fractal dimension
of the scattering function to be equal to $d\simeq 1.84$.
\end{abstract}
\maketitle
\section{Introduction}
Large-scale structures such as vortices, background and tidal currents
are the main constituents in geophysical flows that strongly influence
all the transport and mixing phenomena. Transport of heat and mass plays
a crucial role in dynamics of the ocean and atmosphere. To study the transport
and mixing processes the Lagrangian representation is convenient to adopt.
The motion of a test particle (a tracer) satisfies the differential equation
\begin{equation}
\displaystyle{\frac{d{\bf r}}{dt}={\bf v}({\bf r},t),}
\label{1}
\end{equation}
where ${\bf r}=(x,\,y)$ is the position of the particle and ${\bf v}$
represents the incompressible Eulerian velocity field. It has been
recognized that particle trajectories even in simple flows can be chaotic
in the sense that nearby trajectories separate exponentially in time.
While in a three-dimensional flow it is possible even if the flow is time
independent \cite{A65}, {\it Lagrangian chaos} or {\it chaotic advection}
in a two-dimensional flow may occur only in time-dependent flows
\cite{A84, O89, Z91, C91}.

In this paper we study the dynamics of passive particles (which take on
the velocity of the flow very rapidly and do not influence the flow) in
a simple two-dimensional open flow composed of a fixed point vortex
and a background current with a periodic component. This model is
inspired by a very interesting natural phenomena, {\it topographical
vortices} over mountains, to be found in the ocean and atmosphere
\cite{K83, Z95}. In the deep sea, they have been found in different
regions of the World Ocean with the help of buoys of neutral buoyancy
\cite{Col}. A laboratory prototype of a topographical vortex is a cylindric
anticyclonic vortex to be found by Taylor \cite{T23} in homogeneous fluid over
an underwater obstacle. In the homogeneous ocean the topographical
vortices are cylindric whereas they are of cone-like form in
the stratified ocean.

We adopt, of course, an oversimplified model with a point
vortex in an attempt to catch and quantify the main features
of the tracer dynamics in the flow from the dynamical standpoint.
The fixed point vortex is embedded in a background planar flow
composed of steady and time-dependent components. If the flow satisfies
the incompressibility condition, ${\rm div}{\bf v}=0$, the evolution equation
(\ref{1}) can be written in the Hamiltonian form
\begin{equation}
\begin{array}{l}
\displaystyle{\frac{dx}{dt}=v_x\,(x,\,y,\,t)=-\frac{\partial \Psi}{\partial y}},\\[4mm]
\displaystyle{\frac{dy}{dt}=v_y\,(x,\,y,\,t)=\frac{\partial \Psi}{\partial x}},
\end{array}
\label{2}
\end{equation}
where the streamfunction $\Psi(x,\,y,\,t)$ plays the role of a Hamiltonian.
Thus, the configuration space of an advected particle is a Hamiltonian
phase space of an associated dynamical system. It means that coordinate-space
trajectories of tracers coincide with phase-space trajectories, and
more importantly, the configuration space contains all the typical
structures of Hamiltonian phase space including islands of regular motion,
islands around islands, cantori, stochastic layers, etc. (for a recent review
of Hamiltonian chaos see, for example \cite{Z98}). This prominent non-uniformity of the phase space implies anomalous transport of tracers, including
L{\'e}vy flights, their dynamical trapping and fractal properties, stickiness
to islands boundaries, etc. These phenomena have been studied theoretically
in the field of chaotic advection in a large number of systems (see, for
example, papers \cite{A84, C91, R90, Z88, P99, EA, BB01, JZ} which represent a
small part of relevant publications) and observed in laboratory experiments
\cite{SW93, SK96, ST98, CC87}.

In our model flow, advected particles from the inflow region enter the region
where the fixed point vortex is located and then wash out to the outflow region.
While the tracer trajectories in the inflow and outflow regions, where the
influence of the vortex is negligibly small, are simple as they just follow
the streamlines of the background current, the trajectories in an influence
zone of the vortex, called the {\it mixing region} in the following, are
typically chaotic \cite{BP01}. So, the problem of tracer dynamics in the
open flow we address in this paper is, in fact, a problem of chaotic
scattering \cite{Ott93, JZ, SK96}.

The paper is organized as follows. In Section 2, we introduce the model
streamfunction and the equations of motion of advected particles, consider
the integrable time-independent version of the model flow and prove analytically
transversal intersections of stable and unstable manifolds of the saddle point
producing homoclinic chaos in the time-dependent system. A numerically
simulated image of the homoclinic structure and a typical Poincar{\'e} section
of the particles trajectories in the mixing region provide a general picture of
chaotic advection. In Section 3, we report our main results. We observe and
discuss the phenomenon of stickiness of tracers trajectories to the boundaries
of the vortex core and islands of regular motion which act as {\it dynamical
traps} for tracers. We show that distribution of the dynamical traps over
the space of initial tracer's positions is fractal-like, and the trapping
time distribution demonstrates initially exponential decay followed by a
{\it power-law decay} at the probability distribution function
(PDF) tail with a characteristic exponent
$\gamma\simeq 2$. We compute also dependence of tracer's trapping time on
their initial positions and find it to be a typical fractal scattering function
with an  uncountable number of singularities. The mechanism  of generating the
{\it fractal} is shown to resemble the famous Cantor-set generation with the
Hausdorff fractal dimension of the scattering function to be equal to
$d\simeq 1.84$. Finally, in Section 4, we discuss some possible
applications of the results.

\section{Model flow}
We consider a point fixed vortex embedded in a planar flow of an ideal
incompressible fluid with a stationary and periodic components. The respective
dimensionless streamfunction
\begin{equation}
\displaystyle{\Psi=\ln\sqrt{x^2+y^2}+x\,(\epsilon+\xi\sin\tau)=\Psi_0+\xi\,\Psi_1,}
\label{3}
\end{equation}
generates the Lagrange equations of motion of a passive particle in the flow
\cite{BP01}
\begin{equation}
\begin{array}{l}
\displaystyle{\dot x=-\frac{y}{x^2+y^2}},\\[4mm]
\displaystyle{\dot y=\frac{x}{x^2+y^2}+\epsilon+\xi\,\sin\tau},\\
\end{array}
\label{4}
\end{equation}
where dot denotes differentiation with respect to dimensionless time $\tau$,
$\xi$ and $\epsilon$ are the normalized velocities of a particle in the
stationary and periodic components of the current flowing in the
$y$ direction from the south to the north. Without perturbation,
$\xi=0$, the phase portrait of the dynamical system
(\ref{4}) consists of a collection of finite and infinite trajectories
(streamlines) separated by a loop passing through a saddle point with the
coordinates $(-1/\epsilon;\,0)$. In the polar coordinates
$x=\rho\,\cos{\varphi}$ and $y=\rho\,\sin{\varphi}$,
the unperturbed equations can be solved in quadratures
\begin{equation}
\displaystyle{\epsilon\,d\tau=\biggl[1-\displaystyle{\biggl(\frac{E-\ln\rho}{\epsilon \rho}\biggr)}^{2}\,\biggr]^{-1/2}d\rho},
\label{5}
\end{equation}
where $E=\epsilon\rho\,\cos{\varphi}+\ln\rho$
is the conserved energy. Depending on initial positions and the values of
the control parameters, particles either are trapped inside the separatrix
loop and move along closed streamlines or move along infinite streamlines
outside the loop. In the integrable system, the stable manifold of the saddle
point, along which particles move towards the saddle, and its unstable manifold,
along which they move outward the saddle, coincide. Taking into account the
separatrix value of the energy $E_{s}=-1-\ln\epsilon$,
one can easily derive the travel time for a particle between two points with
the coordinates $(\rho_0,\,\rho)$ nearby the separatrix
\begin{equation}
\displaystyle{T(\rho_0,\,\rho)=\int\limits^{\rho}_{\rho_0}\frac{d\rho'}
{\sqrt{\epsilon^{2}-\displaystyle{\biggl(\frac{E_{s}+\delta-\ln\rho'}{\rho'}\biggr)^{2}}}}},
\label{6}
\end{equation}
where $\delta$ is a small deviation of the energy from its separatrix value.

Under perturbation, the stationary saddle point becomes a saddle periodic
motion which is represented on a Poincar{\'e} section with the dimensionless
period $2\pi$ by a stationary point. In the extended phase space
$(x,\,y,\,\tau)$, there exists a stable (unstable) manifold of the saddle point
with trajectories approaching the saddle periodic motion when $\tau\to\pm\infty$.
Under a typical perturbation, stable and unstable
branches of the unperturbed separatrix loop split and transverse each other
on the Poincar{\'e} section infinitely many times producing a complicated
homoclinic structure. It can be analytically proved with the help of the
Poincar{\'e}-Melnikov integral
\begin{equation}
\displaystyle{I(\alpha)=\int\limits^{\infty}_{-\infty}{\{\Psi_0,\,\Psi_1\}[x_s(\tau-\alpha),\,y_s(\tau-\alpha)]\,d\tau}},
\label{7}
\end{equation}
where $\{\Psi_0,\,\Psi_1\}$  is the Poisson bracket, $x_s$ and $y_s$
are the separatrix solutions of the unperturbed problem parameterized by a
real number $\alpha$. With the streamfunction (\ref{3}) we get the function
\begin{equation}
\displaystyle{I(\alpha)=\sin\alpha\,(\int\limits^{\infty}_{-\infty}d\tau\,\dot x_{s}\cos\tau)-
\cos\alpha\,(\int\limits^{\infty}_{-\infty}d\tau\,\dot x_{s}\sin\tau)},
\label{8}
\end{equation}
that obviously has an infinite number of simple zeroes if
$\dot x_{s}(\tau)\not\equiv 0$. Transversal intersections of stable and unstable
manifolds of a hyperbolic point has been analytically proven with
two-dimensional flows under almost arbitrary nontrivial perturbation
\cite{K95}. 
\begin{figure}[h]
\includegraphics[width=0.485\textwidth, height=0.3\textheight,clip]{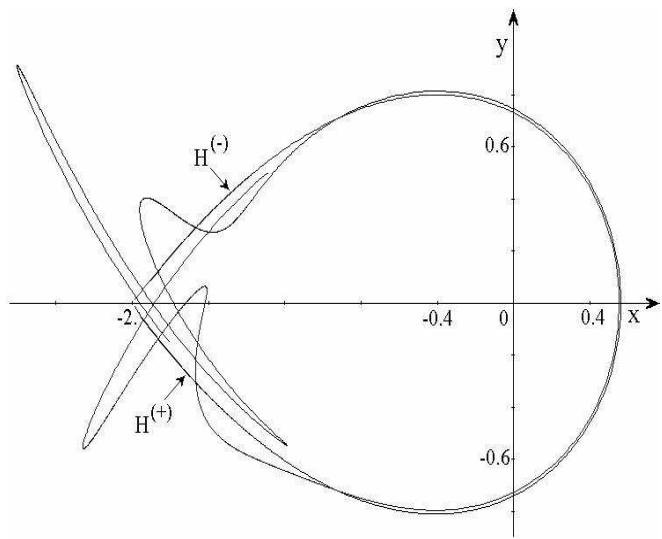}
\caption
{Transversal intersections of stable, $H^{(+)}$, and unstable,
$H^{(-)}$, manifolds of the saddle point for $\epsilon=0.5$ and $\xi=0.01$.}
\label{fig1}
\end{figure}
A simplified image of such intersections is given in FIG.\,1
where a fragment of the Poincar{\'e} section with a small perturbation
$\xi=0.01$ in the time moments $2\pi n\,(n=0,\,1,\,2,\ldots)$ for a bunch
of trajectories, starting from a neighborhood of the saddle point, is shown.
It is a ``seed'' of Hamiltonian chaos that arises under an arbitrary small
perturbation around the unperturbed separatrix loop.

A phase portrait of chaotic advection in our simple flow is shown in FIG.\,2.
The Poincar{\'e} section of passive particle trajectories in FIG.\,2a reveals
a non-uniformity of the phase space that is typical for Hamiltonian systems
with $3/2$ degree of freedom. Fluid around the point vortex placed at
($x=0,\,y=0$) forms a coherent structure, {\it a vortex core}, which is filled
with regular, almost elliptic orbits. Another large coherent structure is seen
to the west from the vortex core, it is filled with regular orbits as well.
A magnification of the northern part of this long island is shown in FIG.\,2b 
which demonstrates
a chain of small islands along its border line. Further magnification
of one of these small islands shown in the inset of FIG.\,2b reveals chains
of smaller islands and so on. A zone with invariant KAM tori embedded in a
stochastic sea, is called the {\it mixing region}. While passive particle
trajectories just follow streamlines outside the mixing region (in the far
downstream and upstream regions) they can be chaotic inside of it.
%
\begin{figure}[h]
\includegraphics[width=0.48\textwidth, height=0.3\textheight,clip]{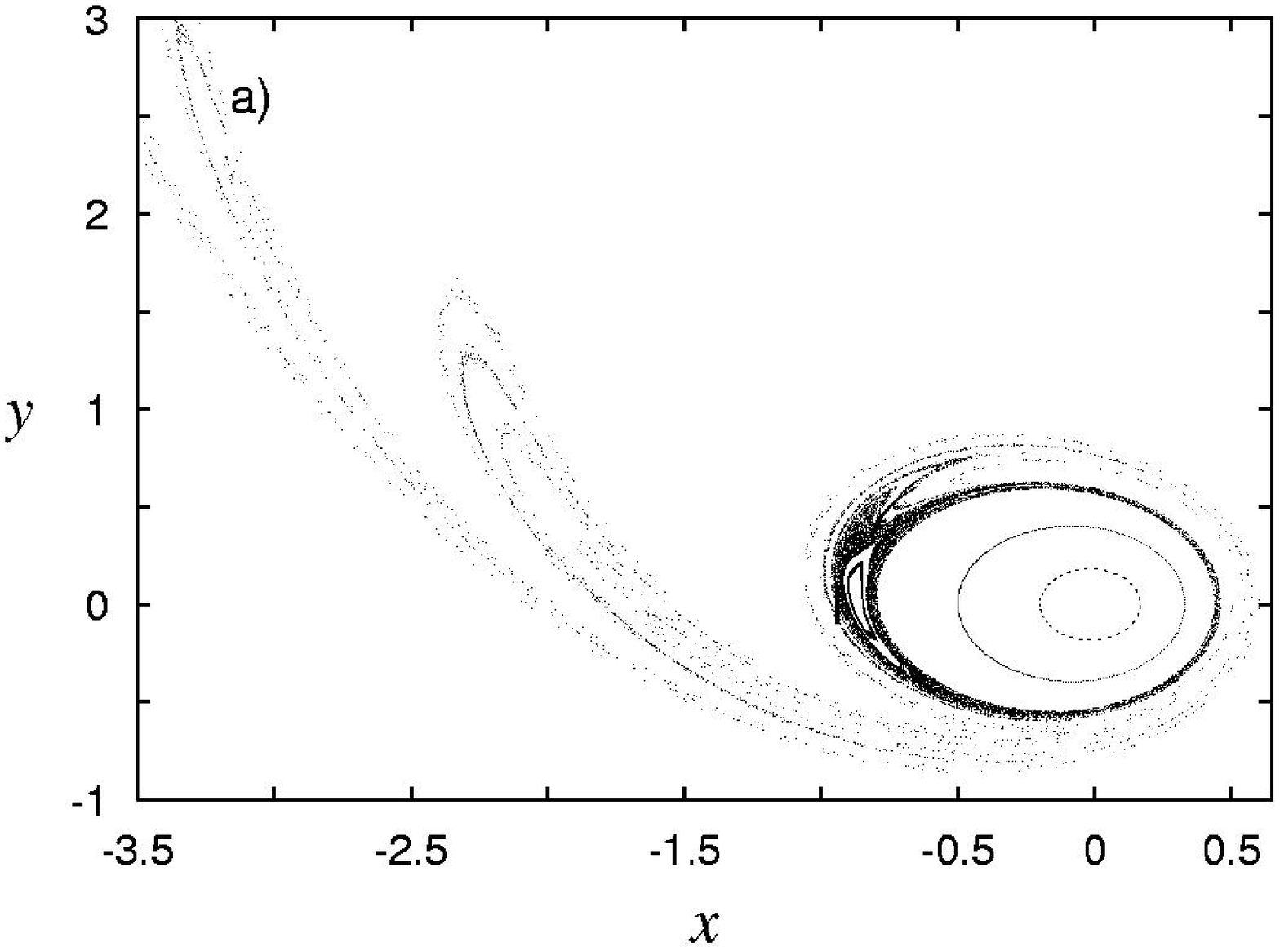}
\includegraphics[width=0.48\textwidth, height=0.3\textheight,clip]{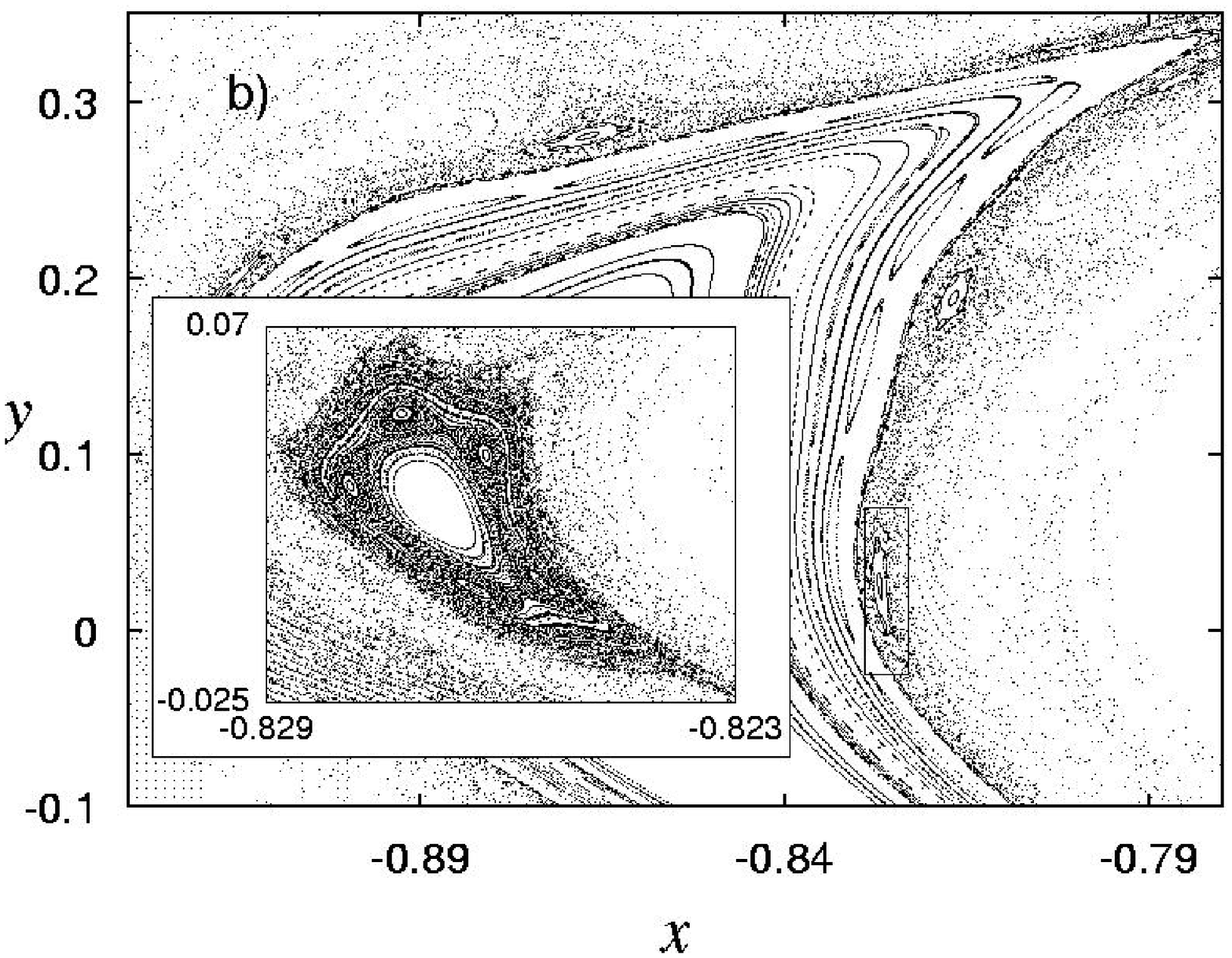}
\caption
{Poincar{\'e} section of the mixing region (a), magnification
of the northern part of the long island (b), and magnification of the
indicated region between the vortex core and the long island in the inset.
Parameters are $\epsilon=0.5$ and $\xi=0.1$.}
\label{fig2}
\end{figure}
%
\begin{figure}[h]
\includegraphics[width=0.485\textwidth,clip]{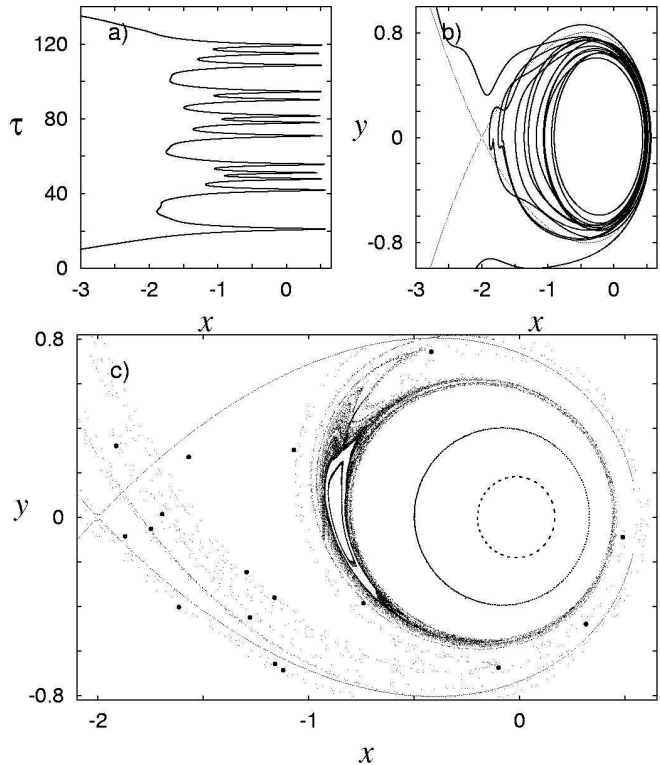}
\caption
{A chaotic wandering of a passive particle initially located
at $(x_0=-4.358533,\,y_0=-6)$ in the mixing region. The time dependence
of the $x$ coordinate (a), the particle's trajectory (b) and the
particle's locations (black circles) on the Poincar{\'e} section
(c) are shown for $\epsilon=0.5$ and $\xi=0.1$. The trapping time
is $T\simeq 115$.}
\label{fig3}
\end{figure}
\section{Dynamical traps and fractals}
Chaotic advection of passive particles in a time-dependent open flow
is a typical problem of chaotic scattering. The southern background current with
a comparatively small alternating component transports particles along
the streamlines into the mixing region, where the particle trajectories
can be very complicated and typically chaotic, and then washes out most of
them to the north where they again follow the streamlines. We show in this
section that in spite of the simplicity of our flow there exists a surprisingly
rich variety of different types of trajectories inside the mixing region.
Particles with small variations of initial positions are trapped there for
trapping times with a broad spectrum extended from a particle transit time
with the average velocity of the current to infinity. The dynamical trapping
cannot be characterized only by the trapping time $T$, the ways by which
advected particles move in the mixing region are very different.

Passive particles are placed outside the mixing region on the line
segment with $y_0=-6$ and different values of $x_{0\,i}$. We compute
the time when the particles reach the line $y=6$ and positions of the
particles, $x_{i}(\tau)$ and $y_{i}(\tau)$. The control parameters
are chosen to be $\epsilon=0.5$ and $\xi=0.1$. In FIG.\,3 we show
the motion of a particle with the initial position
$(x_0=-4.358533,\, y_0=-6)$ trapped in the mixing region for the comparatively
long time $T\simeq 115$. In the upper panel of the figure, the dependence
of the $x$ coordinate of the particle on time $\tau$ (a) and the particle's
trajectory (b) are shown. The Poincar{\'e} plot of the particle positions
at the moments of time $\tau=2\pi n$ (black circles) is shown in FIG.\,3c.
The unperturbed separatrix and the Poincar{\'e} section for a collection
of particles initially placed inside the mixing region are shown for reference.
The particle moves quite uniformly without any preference to visit special zones
in the mixing region.

The second particle to be placed initially close to the first one at the point
$(x_0=-4.358034,\,y_0=-6)$ demonstrates cardinally different behavior shown
in FIG.\,4. It spends in the mixing region much more time, $T\simeq 1581$, than the
first particle does. The figure clearly demonstrates the phenomenon of stickiness
well-known in chaotic Hamiltonian dynamics \cite{D90, ZSW, Z98}. In contrary
to the first particle, the second one spends most of time in the mixing region
nearby the boundaries of the vortex core and the long island. Motion of the
particle in these areas is almost regular with zero maximal Lyapunov exponent.
In Hamiltonian systems, there may exist partially broken invariant
curves (cantori) in a vicinity of island boundaries which prevent particles
to quit the area and act as a kind of permeable barriers. Both these reasons
result in stickiness of the islands boundaries that is seen in FIG.\,4.
%
\begin{figure}[h]
\includegraphics[width=0.485\textwidth,clip]{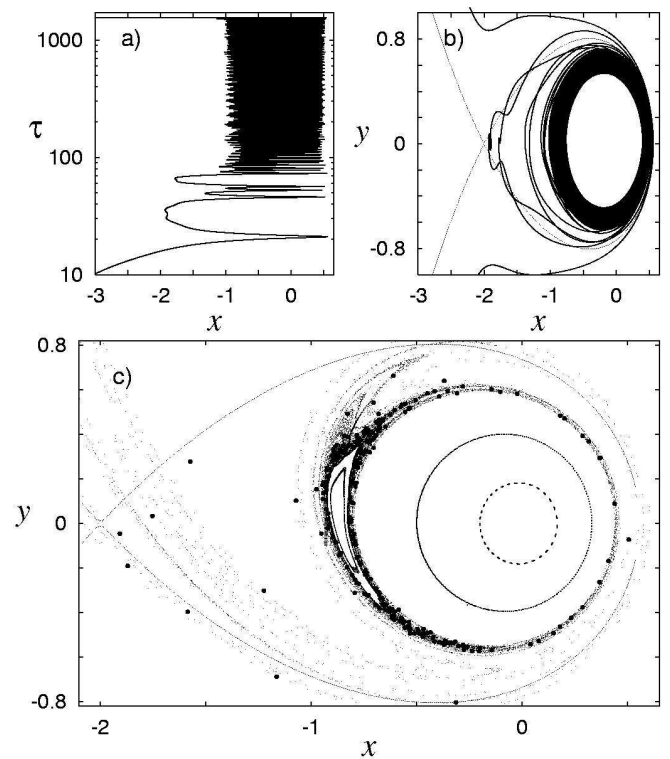}
\caption
{Stickiness of a passive particle initially located
at $(x_0=-4.358034,\,y_0=-6)$ on the boundaries
of the vortex core and the long island. The time dependence
of the $x$ coordinate (a), the particle's trajectory (b) and
the particle's locations (black circles) on the Poincar{\'e}
section (c) are shown for $\epsilon=0.5$ and $\xi=0.1$.
The trapping time $T\simeq 1581$ is very long.}
\label{fig4}
\end{figure}

To study the trapping time distribution we have carried out a number of
numerical experiments. A trapping map referred to initial positions of
$N=22801$ passive tracers, distributed homogeneously on a rectangular
grid $(x_0\in[-5,\,0];\,y_0\in[-6,\,1])$, is shown in FIG.\,5. 
%
\begin{figure*}
\includegraphics[width=0.89\textwidth,height=0.46\textheight,clip]{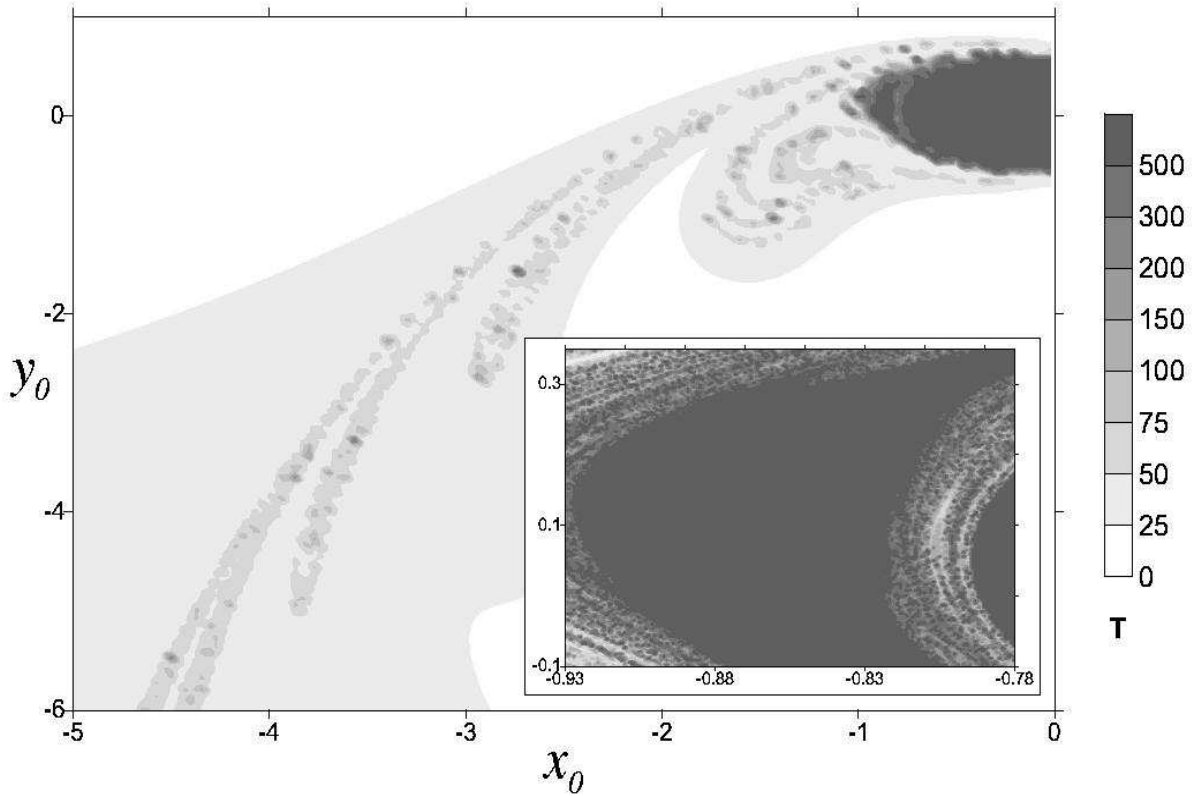}
\caption
{Trapping map showing dependence of the trapping
times in the mixing region $T$ on initial positions of $N=22801$ tracers
distributed on a rectangular grid in the upstream region. The inset
shows a zoom of the fragment of the map.}
\label{fig5}
\end{figure*}
Color modulates
values of the trapping time $T$. The inset shows the respective zone of
this map magnified by the factor of $30$ on the $x$ axis and factor of $17$
on the $y$ axis. A patchiness exists for all the scales revealing a
fractal-like structure of the underlying phase space. Tracers started
initially in the areas of the upstream region, corresponding to black patches
on the map, will be trapped in the mixing region for longer times. The inset
demonstrates that there exist tracers with very close initial positions
whose trapping times differ largely. To quantify this set of
{\it dynamical traps} with a wide range of trapping times we compute
the respective distribution function $N(T)$ for $N=110011$ tracers (positioned
initially on a rectangular grid
$(x_0\in[-4.65,\,-4.35];\,y_0\in[-6.05,\,-5.95])$.  The double logarithmic
plot in FIG.\,6 demonstrates initially exponential decay followed by a
power-law decay at the PDF tail, $N(T)\sim T^{-\gamma}$, with the
characteristic exponent $\gamma\simeq 2$. While a Poissonian distribution is
expected for strong chaotic mixing a power-law dependence indicates the
presence of dynamical traps in the flow. Particles trapped for longer times
give an enhanced contribution to the PDF tail. Different choices of
initial positions of the tracers visiting the mixing region do not
change significantly the trapping-time distribution.
%
\begin{figure}[h]
\includegraphics[width=0.485\textwidth,clip]{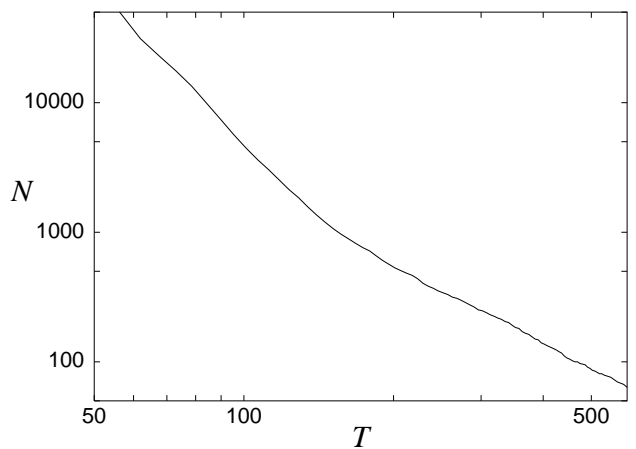}
\caption
{Trapping time distribution with $N=110011$ tracers shown on $\log-\log$ axes.}
\label{fig6}
\end{figure}
\begin{figure*}
\includegraphics[width=0.905\textwidth,height=0.285\textheight,clip]{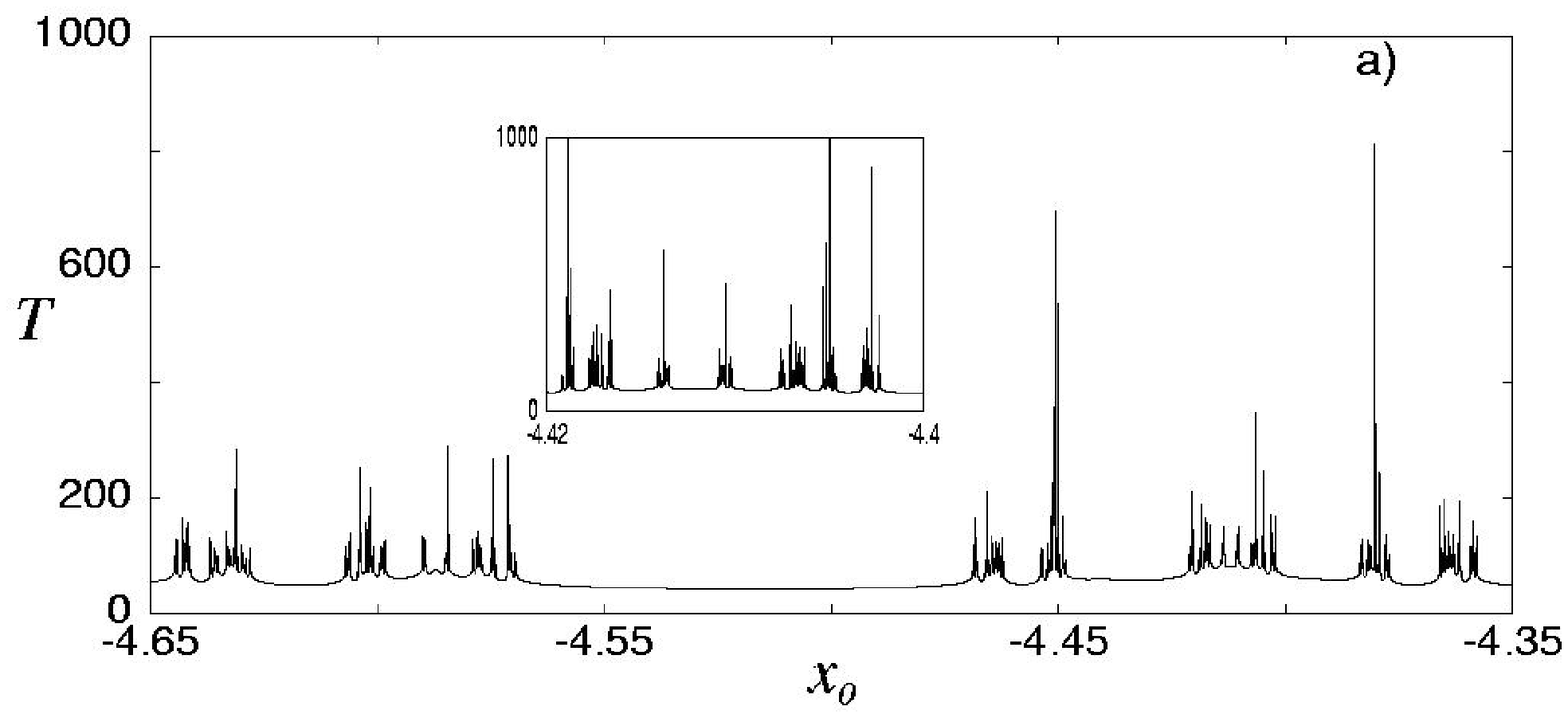}
\includegraphics[width=0.905\textwidth,height=0.285\textheight,clip]{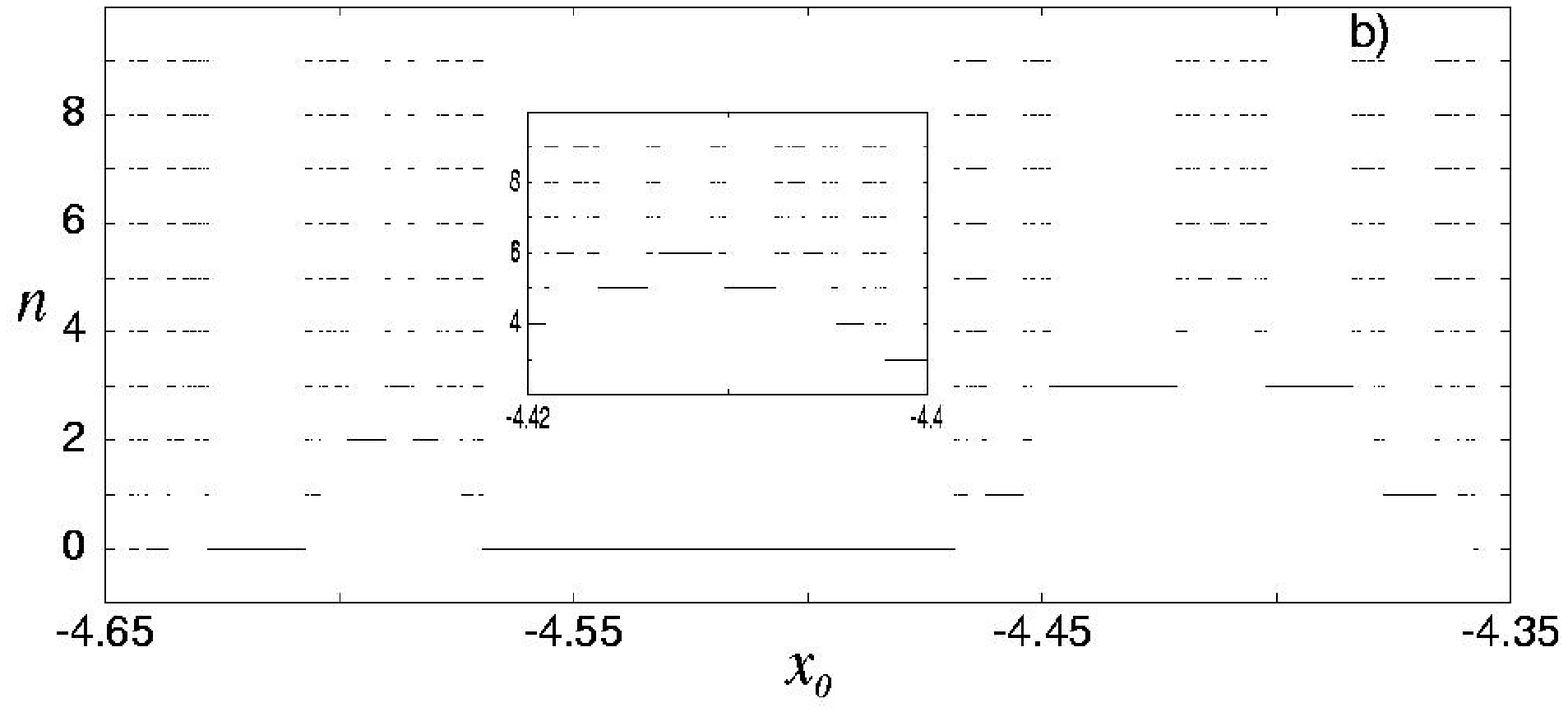}
\caption
{Fractal dependence of the trapping time $T$ on
initial particles positions $x_0$ with the inset showing a 20-fold
magnification of one of the singularity zones (a). Mechanism of
generating the fractal with magnification of a small segment
corresponding to the inset in the top panel (b).}
\label{fig7}
\end{figure*}

The well-known manifestation of chaotic scattering is fractal-like scattering
functions with an uncountable number of singularities
\cite{Ott93, JZ, SK96}. We take the trapping time to be the scattering
function, $T(x_0)$, and compute it carefully for $N=10^4$ particles
distributed initially on the line segment with $y_0=-6$ and
$x_0\in[-4.65,\,-4.35]$. The results in FIG.\,7 demonstrate a typical scattering
function $T(x_0)$ with an uncountable number of singularities which are
unresolved in principle. The inset in FIG.\,7a shows a zoom by the factor
of 20 of one of these singularity zones. Successive magnifications
confirm a self-similarity of the function with increasing values of the
trapping times $T$. FIG.\,4 gives an example of a trajectory with very
long trapping time $T\simeq 1581$. Remind, please, that we measure time in
units of the period of the alternating current. The trapping map in FIG.\,5
provides a two-dimensional image of the fractal structure of the scattering
function $T(x_0,\,y_0)$. As we have shown above, sticky boundaries
of islands of regular motion, embedded in a stochastic sea, act as a kind
of dynamical traps providing all the spectrum of values of the trapping times
up to infinity. Namely these dynamical traps are the ultimate reason for
appearing fractals in our simple model flow.
To give an insight into a mechanism of generating the fractal
we plot in FIG.\,7b segments of the initial string with $10^6$ tracers
which are trapped in the mixing region after $n$ rotations around the
fixed point vortex. After each rotation, a portion of tracers is washed out
in the downstream region with $y\ge 6$. This process resembles the mechanism
of generating the famous Cantor set but is more complicated.

In conclusion, we compute the length of the curve $T(x_0)$ shown in FIG.\,7a
\begin{equation}
\displaystyle{L\,(1/N)=\sum^{N}_{i=1}|T_{i+1}-T_i|,}
\label{9}
\end{equation}
as a function of the inverse number of tracers, $N^{-1}$, distributed initially
on the same line segment with  $y_0=-6$ and $x_0\in[-4.75,-4.25]$ in the
upstream region of the flow. It is a decreasing (in average)
function with a negative slope whose double lagarithmic plot is 
presented in FIG.\,8. The Hausdorff dimension of the 
fractal-like scattering function is computed to be 
$d\simeq 1.84$.
%
\begin{figure}[h]
\includegraphics[width=0.485\textwidth,clip]{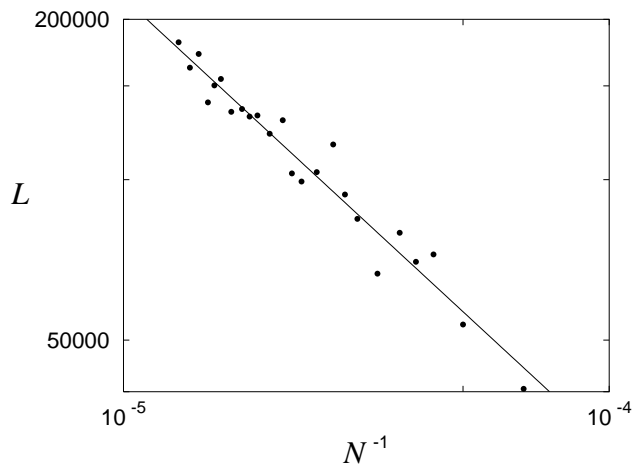}
\caption
{Length of the fractal curve $T(x_0)$ versus
the inverse number of tracers $N^{-1}$ distributed initially on the same
line segment in the upstream region. The double logarithmic plot.}
\label{fig8}
\end{figure}
\section{Conclusion and discussion}

We have shown that chaotic advection of passive particles by the open flow,
composed of a fixed point vortex and a background current with a time-dependent
component, has a fractal nature. Appearance of fractals and dynamical traps in
geophysical flows with topographical vortices should fluence strongly transport
and mixing of heat and mass. Instead of homogeneous mixing, one expects a
highly-structural, hierarchical fractal distribution of advected particles
that is of great interest in oceanography and atmospheric sciences. It should
be emphasized that the fractals and dynamical traps are not specific for our
particularly simple model. They are typical in low-dimensional Hamiltonian
systems and should exist in more realistic models with elliptic topographical
vortices and different boundary conditions considered, for example, 
in \cite{KK}. 

If tracers are not passive but active (chemical reagents or 
biological species), their activity is superimposed on chaotic mixing
in open background flows. Typically, there exists a local coupling 
between a spatial character of the activity and the spatial character 
of the fluid with coherent structures, dynamical traps and fractals. Say in
chemistry, molecules will react if they are in a close contact with
each other. So, the imperfect mixing should largely increases the 
chemical or biological activity in some regions of the flow
and suppress it in other regions. In marine biology, the paradoxical
observational fact of coexisting thousands of phytoplankton species
in spite of a limited number of resources for which they compete has
been known for a long time \cite{H61}. Chaotic mixing of active 
particles in open nonstationary flows in the ocean with small-scale
spatial inhomogeneities and fractal patterns can provide a natural
explanation of the plankton paradox \cite{K00}. Homogeneous and 
complete mixing leads to an equilibrium state in the system of
competing species resulting in the survival of the most fitted
species, whereas chaotic mixing generates, typically, a persistent
nonequilibrium state in the system providing the coexistence of much
more species than the number of ecological niches they occupy. 

\section{Acknowledgments}
This work is supported by the Russian Federal Program ``World Ocean''
in the framework of the Project ``Modelling variability of hydrophysical
fields'' and by the Russian Foundation for Basic Research under
project numbers 02-02-17796 and 02-02-06841. We gratefully acknowledge
useful discussions with V.F. Kozlov.

\pagebreak
\end{document}